# Syntheses of a heterogeneous catalysts using vapor phase self-assembly.

Vladimir Burtman

*University of Utah, Physics Department, 115 S. 1400 E. Suite 201, Salt Lake City, Utah 84112-0830, USA*

**The first vapor phase self-assembled heterocatalytic structure on surface is build from cobalt 5,10,15,20 tetrakis (4-aminophenyl)-21H, 23H-porphine through imide bonds. This heterogeneous catalyst possesses essential thermal and temporal stability along with catalytic activity in oxidation processes of organic substrates.**

The heterogenisation of homogenous catalysts is rapidly becoming an important area of chemistry.[1] It is hoped that advantages of homogeneous (i.e. high reaction activity) and heterogeneous (easy separation of products and reagents from catalysts) can be combined into one system.

In many cases the applications of surface-modified materials have not been successful because: 1) short lifetime of catalysts caused by leaching of reactive groups, 2) steric effects of the matrix, and 3) unhomogeneity of reactive centers.[2] In this work we present a novel way of heterogenisation of a wide range of catalysts (organic, metallorganic, inorganic) through vapor phase self-assembly in order to obtain materials with high thermal and temporal stability properties and good catalytic activity.

A novel method enable the introduction of modified homogenic catalyst as an active catalytic layer in metallorganic heterostructure by the ultra high vacuum (UHV) vapor phase self-assembling (VP-SAM) method, followed by the formation of covalent bonds between interfaces. The possibility of this synthetic route was shown on the example of naphthalene diimide organic superlattices for optoelectronic applications[3]. In present contribution we demonstrate UHV vapor-phase self-assembly synthesis of heterogenic catalysis. We demonstrate VP-SAM approach to heterogenisation catalysis on example of a catalytic organic heterostructure containing the layer of cobalt 5,10,15,20 tetrakis (4-aminophenyl)-21H, 23H-porphine (CoTAPP). Incorporation of such porphine units into supramolecular structure is favorable for following reasons. CoTAPP molecules are robust, rigid and highly symmetrical and thus serve as desirable, slab-like building blocks for synthesizing microporous materials with large cavities and channels that have potential catalytic activity.[4]

CoTAPP was synthesis according to of Adler and Dutta-Gupta procedures.[7] A template layer is firstly deposited on oxide surfaces, such as $Si/SiO_2$ and glass (Fig. 1, **A** step *i*), by reaction of 3-aminopropyl-trimethoxysilane (**A**, acylation route) with surface exposing free amine functionality toward the next synthetic step (Fig. 1, **A** step *i*, substrate temperature 100 °C, total pressure 0.05 Torr) [8].

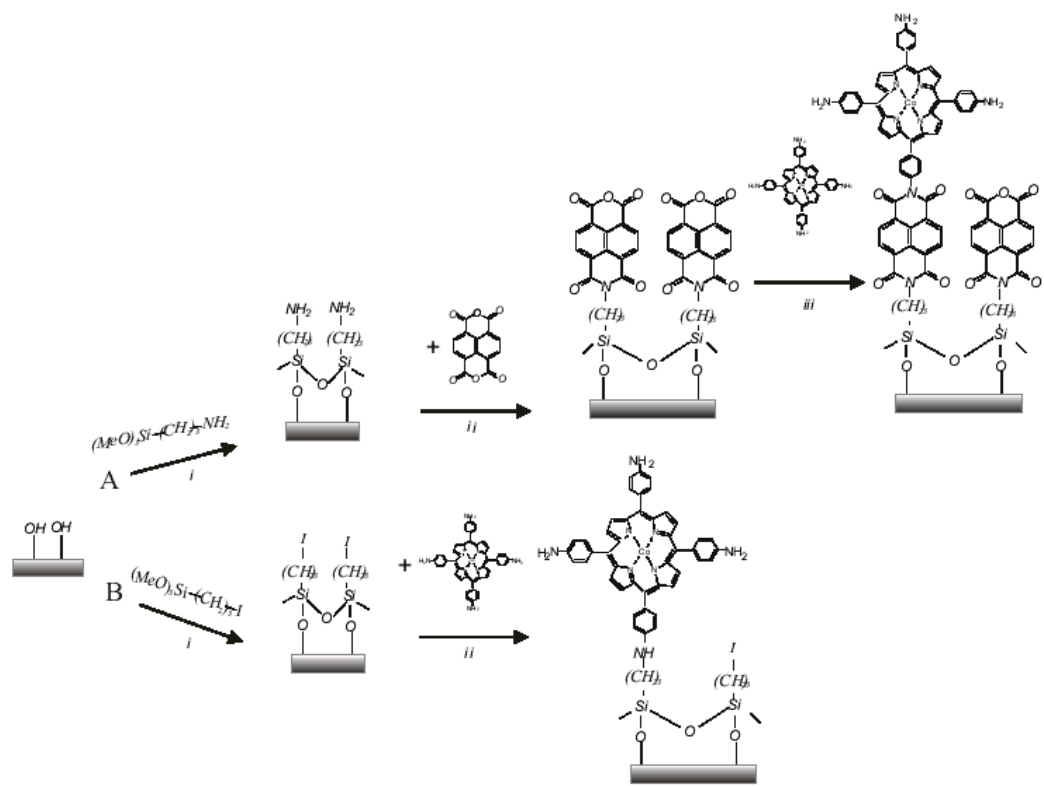

Scheme I Vapor phase SAM of Co-porphyrin

Then an alkylamine-containing surface is hit with a pulse of 1,4,5,8-naphthalene-tetracarboxylic-dianhydride (NTCDA) precursor (Fig. 1, **A** step *ii*), forming imide linkages[3]. Then a pulse of a vaporized CoTAPP was linked to tail 1,8-naphthalic anhydride groups of spacer layer (Fig. 1, **A** step *iii*) following by formation of imide bonds. Alternatively the CoTAPP containing heterostructure was achieved through amin-alkylated route **B**. A template layer deposited by reaction of 3-iodopropyl-trimethoxysilane (**B**, alkylated route) with surface exposing free halogen functionality toward the CoTAPP synthetic step (Fig. 1, **B** step *i,* same conditions as in **A** ,step *i*). Then an iodo-containing surface is hit with a CoTAPP precursor (Fig. 1, **B** step *ii*), forming secondary amine.

CoTAPP was evaporate at T = 200 °C, at total pressure of $10^{-3}$ Torr, transported to reaction zone by Ar carrier gas and then reacted with anhydride surface at T = 220 °C with following formation of imide bridge. Excess of CoTAPP and water (co-product) were evacuated by vacuum from reactive zone.

Model compound CoTAPP-1,8-naphtaloanhydride (CoTAPP-NA) was achieved by the modified synthesis reported in ref. 9. Model compound was studied in order to examine an enhanced thermal and temporal stability of CoTAPP-containing films. Direct scanning calorimetry (DSC) of CoTAPP-NA exhibit the growth of decomposition temperature from 340 °C for CoTAPP to 550 °C for CoTAPP-NA. The thermal behavior of model compound mimics the enchanted stability of CoTAPP assembled heterostructures.

The resulting thin films were studied ex situ by UV-Vis and FTIR spectroscopy, spectroscopic ellipsometry, evaluation of an absolute number of free amino-groups on building interface and estimation of contact angle.

UV-Visible spectroscopy of CoTAPP obtained by **A** route contained heterostructure revealing a red-shift from 433 nm to 435 nm which is characteristic of porphyrin 'Soret band'[10] and corresponding to the formation of imide bridges.[11] This shift is consonance with the red-shift trend of Soret peak in porphyrins containing electron withdrawing groups. In addition, the characteristic naphthalene monomer bands of 354,5 and 383,0 nm also appeared in absorption spectra. UV-Visible spectroscopy of CoTAPP assembled through secondary amin formation (**B** route) reveal an appearance of characteristic aromatic bands at 256nm and 370 nm in addition to weak 'Soret band'. The appearance of those bands is additional evidence of aromatic-containing structure formation.

The surface condensation reaction between alkylamine and NTCDA (**A** route) was studied by FTIR measurements of alkyl-amine monolayer self-assembled on Au/SiO2/Si substrate. The alkyl-amine peak at 3250 cm$^{-1}$ disappeared and the imide peak at 1658 cm$^{-1}$ appears along with unreacted anhydride group peak at 1725 cm$^{-1}$ after reaction with NTCDA in MLE-derived films on gold. Following attachment of CoTAPP and formation of second imide bond lead to formation of two imide bands at 1664 cm$^{-1}$ and 1657 cm$^{-1}$ and spread peak of aromatic aminogroups at 3210 cm$^{-1}$

Absolute surface density of amine groups of the (aminopropyl) surface was determined using 4-nitrobenzoaldehyde titration of amino- coated substrates according to procedure described in the ref 13. Firstly it was verified the fabrication of first template layer and get the absolute density of 3.75/100 Å$^2$. However for **A** synthetic route this procedure can not be used to determine the total amine density of CoTAPP on surface. It is clear that in addition to aromatic NH$_2$ groups of CoTAPP there are also unreacted NH$_2$ groups of coupling layer. In order to eliminate the coupling factor we study **B** synthetic route, where CoTAPP was attached directly on 3-iodopropyltrimetoxysilane coupling agent. In last case only aminogroups of CoTAPP are present on the surface and thus the unambiguous measurement of absolute amine surface density is possible. The absolute density reaches 4.84/ 100 Å. Dividing the density number on 3 free $NH_2$ groups per CoTAPP molecule we able to estimate the number of active catalytic centers in heterostructure. Finally the measurement of absolute amine surface density evident to fabrication of CoTAPP-containing heterostructure.

Spectroscopic ellipsometry of CoTAPP heterostructure reveal the 20 Å thickness of CoTAPP monolayer. Contact angle was measured after every deposition step. Contact angles were: 60°, 90°, 55° for **A** *i -iii* steps respectively and 70°, 60° for **B** *i -ii* steps.

The catalytic activity of hetero-assembled macromolecular structure to oxidation processes of some organic substrates in the air was studied by GS and GS/MS techniques. In a typical experiment 10 mg ($\approx 5\times10^{-5}$ mol) of substrate was heated in a high open tube in presence with about $10^{-9}$ mol of CoTAPP catalysis[12] without solvents (substrate/catalytic ratio is $\approx 5\times10^4$). CoTAPP heterostructure consists on planar building blocks, enabling direct oxygen access to catalytic centers, thus steric factors do not decrease catalytic activity. We summarize these results of catalytic activity of CoTAPP (**B** route) VP-SAM films in Table 1.

Table 1. Catalytic activity of CoTAPP VP-SAM films.

| Substrate | Product | Temperature, °C | Time, h | Yield, % | Without catalyst, % |
|---|---|---|---|---|---|
| Xanthene* | Xanthone | 140 | 3 | 34 | 12 |
| Fluorene | Fluorenone | 140 | 12 | 16 | - |
| 9-Anthrone* | 9,10 Anthraquinone | 140 | 3 | 30 | 18 |
| 2,6 Dimethyl-naphtalene | *i* 2,6-(Dihedroxymethyl) naphtalene | 140 | 12 | 25 | 10 |
|  | *ii* 6 Methyl 2-(Hydroxymethyl naphatlene |  |  | 7 | 2 |
| Diphenyl metane | Benzophenone | 140 | 12 | 7 | - |

*after 12 hours full conversion was obtained

The of catalytic activity of CoTAPP assembled by **A** route was almost the same +/- 10 %.

The effect of immobilizing the metalloporphyrin upon catalyses activity, i.e. comparison with homogeneous reactions was also studied on example of possible foxidation.[note 1] No traces of 9-fluorenone were detected.

Finally the straightforward VP-SAM method was demonstrated on the incorporation of metallorganic CoTAPP catalysts in imide-bridged organic heterostructure. The chemical, temporal and thermal stability of these organic structures, along with good catalytic activity toward oxidation processes (70-220 cycles), were proved to be useful in catalytic applications. Catalytic activity remains the same within 7 oxidation experiments. These heterogeneous catalysts are ready for use in the ambient atmosphere. We showed that in addition to previously demonstrated molecular electronic applications of the VP-SAM method,[3] there is useful potential of this synthetic and instrumental method in heterogenius catalysis. We demonstrate that aside of imide-based bonds[3] secondary amine bonds can be also achieved by VP-SAM route. Prolonged thermal, operational stability of the resulting catalyst and process flexibility is favorable for a number of practical catalytic applications.

**Notes and references**

Note [1] 1,8gr of fluorene($1\times10^{-2}$mol) was refluxed together with 5 mg($7.8\times10^{-6}$mol) of CoTAPP. (substrate/catalyst relation is 1300:1) in mixture of 1,2-dichlorobenzene and DMF 20ml (5:1) 24hours.